\documentclass[nofootinbib,twocolumn,prd,preprintnumbers,superscriptaddress]{revtex4-1}

\usepackage{amsfonts}
\usepackage{graphicx}
\usepackage[colorlinks=true,linkcolor=blue,citecolor=red,urlcolor=blue]{hyperref}
\usepackage{color}
\usepackage[dvipsnames]{xcolor}
\usepackage{amsmath,bm}
\usepackage{cases}
\usepackage{braket}
\usepackage[T1]{fontenc}
\usepackage{mathrsfs}

\begin{document}
\title{Joint estimation of the cosmological model and  the mass and redshift distributions of the binary black hole population  with Einstein Telescope}

\author{Matteo Califano}
\email{matteo.califano@unina.it}
\affiliation{Scuola Superiore Meridionale, Largo San Marcellino 10, 80138 Napoli, Italy}
\affiliation{Istituto Nazionale di Fisica Nucleare (INFN), Sezione di Napoli, Via Cinthia 21, 80126 Napoli, Italy}

\author{Ivan De Martino}
\email{ivan.demartino@usal.es}
\affiliation{{Departamento de F\'isica Fundamental, Universidad de Salamanca, Plaza de la Merced, s/n, E-37008 Salamanca, Spain}}
\affiliation{{Instituto Universitario de Física Fundamental y Matemáticas, Universidad de Salamanca, Plaza de la Merced, s/n, E-37008 Salamanca, Spain}}

\author{Daniele Vernieri}
\email{daniele.vernieri@unina.it}
\affiliation{Dipartimento di Fisica ``E. Pancini'', Università di Napoli ``Federico II”, Via Cinthia 21, 80126 Napoli, Italy}
\affiliation{Scuola Superiore Meridionale, Largo San Marcellino 10, 80138 Napoli, Italy}
\affiliation{Istituto Nazionale di Fisica Nucleare (INFN), Sezione di Napoli, Via Cinthia 21, 80126 Napoli, Italy}

\begin{abstract}
We investigate the capability of constraining the mass and redshift distributions of binary black hole systems jointly with the underlying cosmological model using one year of observations of the Einstein Telescope. {To this aim, we fixed the underlying cosmological model to a flat $\Lambda$CDM model, then we considered the mass distribution given by a smoothed power law, and the redshift distributions given by the Madau-Dickinson model. We built mock catalogs with different SNR thresholds, and finally inferred astrophysical and cosmological parameters jointly adopting a hierarchical Bayesian framework. We found that as the SNR threshold decreases, the precision on the matter density parameter $\Omega_{m,0}$ and the Hubble constant $H_0$, improves significantly due to the increased number of detectable events at high redshift. However, degeneracies between cosmological and astrophysical parameters exist and evolve with the SNR threshold. Finally, we showed that one year of observations will serve to reconstruct the mass distribution with its features. Conversely, the redshift distribution will be poorly constrained and will need more observations to improve.}  
\end{abstract}

\preprint{ET-0077A-25}
\maketitle

\section{Introduction}
Gravitational waves (GWs) emitted by inspiraling binary systems serve as precise cosmological distance indicators, generally known as \emph{standard sirens}~\cite{Schutz:1986gp,Holz:2005df,Palmese:2025zku}. This term draws an analogy to \emph{standard candles} emitting in the electromagnetic spectrum, such as Cepheids and Type Ia supernovae (SNeIa), which provide distance measurements based on their intrinsic luminosity. While \textit{standard candles} require a calibration ladder to estimate distances, \textit{standard sirens} offer a direct measurement of distance based only on the gravitational waveform.

When two compact objects, such as binary black holes (BBH) or binary neutron stars (BNS), merge, they emit GWs \cite{Dirkes:2018nkq,LIGOScientific:2016aoc, LIGOScientific:2017vwq} that carry detailed information about the intrinsic properties of the system such as masses, spins, and the distance to the observer. By accurately modeling the evolution of the gravitational waveform amplitude and frequency over time, astronomers can directly determine the luminosity distance to the source without the need for additional calibration steps. This makes \textit{standard sirens} robust and independent tools for cosmological tests, unaffected by uncertainties and systematics that plague traditional distance indicators.

The direct distance measurement capability of \textit{standard sirens} enables them to tackle fundamental cosmological questions, such as determining the Hubble constant ($H_0$), investigating the universe’s expansion history, and constraining the properties of Dark Energy (DE) \cite{DiValentino:2025sru}. 
One of the main difficulties is the degeneracy between the source's mass and redshift, which complicates the accurate estimation of distances unless an independent redshift measurement is available. This limitation was pointed out with the detection of the event GW170817, when the identification of an electromagnetic counterpart was crucial to breaking this degeneracy \cite{LIGOScientific:2017ync,LIGO_H0_2017}. 
The detection of a kilonova and a gamma-ray burst (GRB) from a BNS merger enabled an independent and precise measurement of $H_0 = 70.0^{+12.0}_{-8.0}\, \text{Mpc}\,\text{km}^{-1}\,\text{s}^{-1}$ \cite{LIGO_H0_2017}. Such events, whose redshift can be measured independently, are referred to as \emph{bright sirens}~\cite{Holz:2005df,Dalal:2006qt,Nissanke:2009kt}.
However, for most of the GW events, electromagnetic counterparts are not observed, classifying these events as \emph{dark sirens}.
In the absence of electromagnetic counterparts, alternative methods are required to estimate the redshift of the GW sources. These methods include translating the redshift mass distribution of the GW signal to the source-frame mass distribution using astrophysical models that also include pair instability supernovae (PISN) to constrain the redshift mass distribution ~\cite{Bond:1984sn,Belczynski:2017gds,Woosley:2016hmi,Spera:2017fyx,vanSon:2020zbk,Mastrogiovanni:2021wsd,Mukherjee:2021rtw,Ezquiaga2022,Renzo:2022rnk,Mastrogiovanni:2023emh,Hendriks:2023yrw}. Another approach involves statistical associations of GW events with galaxy catalogs to estimate the source's redshift~\cite{Schutz:1986gp,DelPozzo:2011vcw,Nishizawa:2016ood, Mukherjee:2022afz}. Additionally, spatial clustering of GW sources with galaxies can be used to infer the redshift, using the distribution of GWs relative to galaxy distributions~\cite{Mukherjee:2019wcg,Mukherjee:2020hyn}. 
Another approach utilizes prior knowledge of the redshift probability distribution of merging sources, derived from intrinsic merger rates estimated by population synthesis models~\cite{Ding:2018zrk,Ye:2021klk}. In the case of BNS mergers, the imprinted tidal interactions in the GW signal can also serve to estimate the redshift~\cite{Messenger:2011gi,Chatterjee:2021xrm,Ghosh:2022muc,Ghosh:2024cwc}.

Recently, using BBH redshifted masses, the LVK collaboration
simultaneously infers the source mass distribution and $H_0$~\cite{LIGO_H0_2021} bounding the Hubble constant to $H_0 = 68^{+12}_{-8}\, \text{Mpc}\,\text{km}^{-1}\,\text{s}^{-1}$ at 68\% of the credible level {in the case the redshift is obtained from galaxy catalogs}. This result was also improved by looking at a putative host galaxy leading to $H_0 = 68^{+8}_{-6}\, \text{Mpc}\,\text{km}^{-1}\,\text{s}^{-1}$. 
As GW detection capabilities evolve, the role of \textit{dark sirens} in cosmological studies will become increasingly prominent.
In \cite{Muttoni2023}, it has been shown that an accuracy less than 1\% $H_0$ is achievable using observations of \textit{ dark sirens} with a signal-to-noise ratio greater than 300.
Furthermore, in \cite{Ghosh:2022muc} the equation of state and the Hubble constant are simultaneously inferred using the tidal deformation of the BNS, achieving an accuracy of 5\% in $H_0$.

Here we want to exploit the dark sirens observation in light of the third-generation (3G) GW detectors, such as the Einstein Telescope (ET) \cite{Punturo:2010zz,Branchesi:2023mws,Abac:2025saz}, to estimate their capability to advance our understanding of cosmological models and the expansion history of the universe. In Section \ref{sec:model_data}, we summarize the underlying theoretical model, the hierarchical Bayesian framework that we will use in our approach, and the generation of the mock data. In Section \ref{sec:results} we show our results and discuss assumptions and limitations.
Finally, in Section \ref{sec:conclusions} we give our final discussion and conclusions.

\section{Theoretical model, mock data, and data analysis. }\label{sec:model_data}

We want to investigate the capability of ET to constrain the cosmological model and the astrophysical population distribution jointly. To this aim, we first must generate mock catalogs of the observations that ET will provide and then we have to carry out a forecast analysis to estimate the accuracy down to which ET will estimate the parameters of interest. Therefore, in this section, we will start to describe the underlying cosmological and astrophysical models that we will use as setup for the generation of the mock catalogs. Then we will move on to explaining how we will generate the mock catalogs, and finally we will introduce the hierarchical Bayesian framework employed in the forecast analysis.

\subsection{Defining the parameter space}

Given a set of $N_{obs}$ observations of GWs events, we label with $\theta_i$ the parameters inherent to the gravitational waveform while $\bm \Lambda$ include the astrophysical population and cosmological parameters, namely  $ \bm\Lambda = \{ \bm \Lambda_{astro}, \bm \Lambda_{cosmo}\}$. More specifically, the single-event parameters that are measured from the gravitational waveform are the following 
\begin{equation}
\bm{\theta}  =\{ m_1^d,m_2^d,d_L, \iota, \alpha,\delta, \psi, t_c,\Phi_c,\chi_1,\chi_2 \}\,,
\end{equation}
where $m_1^d$ and $m_2^d$ are the masses of binary system in the detector frame\footnote{We label the masses in the detector frame with the apex $^d$, otherwise we use just $m_1$ and $m_2$.}; $d_L$ is the luminosity distance to the source; $\iota$ is the inclination angle of the plane of the binary system with respect to the line of sight; $\alpha$ and $\delta$ are right ascension and declination, respectively; $\psi$ is the polarization angle; $t_c$ is the time of coalescence; $\Phi_c$ is the phase at coalescence, and $\chi_i$ are the dimensionless spin of the the two members of the binary system. Recently, it has been shown that the spin distributions are difficult to reconstruct even in analysis where the cosmology is fixed \cite{DeRenzis:2024dvx}. {Nevertheless, it has been shown that joint mass--spin models can (slightly) improve cosmological measurements \cite{Tong:2025xvd}.} Since our aim is to investigate the capability of fitting at the same time the cosmological model and the astrophysical population distributions of masses and redshift, we decided to focus on the following subset of parameters: $\bm{\theta} =\{ m_1^d,m_2^d,d_L\}$. 
Furthermore, we divided the population hyperparameters as 
$\bm \Lambda_{astro} = \{\bm \Lambda_m, \bm \Lambda_z\}$, where
$\bm \Lambda_m = \{m_{min}, m_{max}, \alpha,\beta, \mu_g, \sigma_g, \lambda_g, \delta_m\}$ denotes the parameters that describe the mass distribution,  and $\Lambda_z = \{ \gamma, \kappa, z_p\}$ denotes the parameters that enter in the merger rate density.

\subsection{Underlying Astrophysical and Cosmological models}\label{sec:theory}

To simulate a catalog of BBH mergers which will be detectable by ET, we need to set the fiducial astrophysical and cosmological models. 

The astrophysical model depend on the mass and redshift distributions. Following \cite{Fishbach:2018edt}, we assume that $\bm \Lambda_m$ does not evolve with redshift, thus the mass distribution can be written as \cite{Talbot:2018cva}:
\begin{equation}
    \label{eq:mass distrib}
    p(m_1 , m_2 | \bm \Lambda_m)= p(m_1 | \bm \Lambda_m)p(m_2 |m_1, \bm\Lambda_m)\,,
\end{equation}
where 
\begin{align}
 p(m_1 | \bm \Lambda_m) =&(1-\lambda_g)\mathcal{P}(m_1|m_{min}, m_{max}, -\alpha)\nonumber\\
 &+ \lambda_g\mathcal{G}(m_1|\mu_g , \sigma_g)\,. \label{eq:mass_distribution1}
\end{align}
and
\begin{equation}\label{eq:mass_distribution2}
    p(m_2 |m_1,\bm \Lambda_m)=\mathcal{P}(m_2|m_{min}, m_{1}, -\beta)\,.
\end{equation}
The function $\mathcal{P}(x|x_{min}, x_{max}, \xi)$ is a power-law distribution with $\xi$ as slope parameter defined in the interval $\left[ x_{min},x_{max}\right]$;  $\mathcal{G}(x|\mu , \sigma)$ is a Gaussian distribution with mean and standard deviation given by $\mu$ and $\sigma$, respectively. Moreover, the parameter $\lambda_g$ specify the fraction of events that follows the Gaussian distribution. 
Additionally, we add to equation \eqref{eq:mass distrib} a smoothing factor when $m = m_{min}$ in order to avoid  a hard cut-oﬀ, namely \cite{Talbot:2018cva,LIGOScientific:2020kqk}
\begin{equation}
\begin{aligned}
        p(m_1 , m_2 | \bm \Lambda_m )=& p(m_1 | \bm \Lambda_m)p(m_2 |m_1,\bm \Lambda_m)\times\\
        &\times\Sigma(m_1|m_{min}, \delta_m)\Sigma (m_2|m_{min}, \delta_m).
        \end{aligned}
\end{equation}
The smoothing function $\Sigma(m|m_{min}, \delta_m)$ is defined as follows 
\begin{equation}
\begin{aligned}
        &\Sigma(m|m_{min}, \delta_m)=\\
    &=\begin{cases}
        0   &\text{for } m>m_{min}\\
        f(m- m_{min},\delta_m) &\text{for  } m_{min}\leq m < m_{min}+\delta_m\\
        1 &\text{for  } m \geq m_{min}+\delta_m
    \end{cases}\,,
\end{aligned}
\end{equation}
where 
\begin{equation}
    f(m,\delta)=\left[1 +\exp{\left( \frac{\delta}{m}+\frac{\delta}{m-\delta_m} \right)}  \right]^{-1}.
\end{equation}
Finally, we adopt the Madau-Dickinson model to describe the redshift distribution  \cite{Madau:2014bja}:
\begin{equation}
    \label{eq:redsf distrib}
    \psi(z|\Lambda_z) = R_0 \left[ 1 + (1+ z_p)^{-\gamma-\kappa}\right]\frac{(1+z)^\gamma}{1 + \left[\frac{(1+z)}{(1+ z_p)}\right]^{\gamma+\kappa}}
\end{equation}
where $z_p$ labels the peak of the star formation rate, and $\gamma$ and $\kappa$ are constants that parameterize the behavior at low ($z<z_p$) and high redshift ($z>z_p$), respectively. Finally, $R_0$ is the local rate of GW events, and usually expressed in $\text{Gpc}^{-3}\ \text{yr}^{-1}$.
It is worth noting that in our analysis, we neglected any redshift dependence in the mass distribution. Although there is currently no evidence for such dependence in the redshift range explored by the LVK Collaboration \cite{Lalleman:2025xcs}, {the debate is still open \cite{Rinaldi:2023bbd}}, and this assumption may not hold for third-generation detectors \cite{Fishbach:2021yvy,vanSon:2021zpk,Belczynski:2022wky}.

The underlying cosmological models is instead set to a flat $\Lambda$CDM model. Therefore, the luminosity distance is
\begin{align}
\label{eq: d_L}
    d_L(z)=\frac{c(1+z)}{H_0} \int^z  \frac{dz'}{E(z')}\, ,
\end{align}
where $E(z)=\sqrt{\Omega_{m,0}(1+z)^3 +\Omega_{\Lambda,0}}$, and $\Omega_{m,0}$ and $\Omega_{\Lambda,0}$ are the matter and cosmological constant critical densities measured at $z=0$, respectively.
The cosmological parameters enter in the distance-redshift relation, hence the redshift estimation depends on the cosmological model affecting the link between the detector frame to source frame for the masses, $m^d =(1+z)m$.
Furthermore, to estimate the redshift we have to model the source mass and the redshift distributions  \cite{Mancarella:2021ecn,Mastrogiovanni:2021wsd,LIGOScientific:2021aug}. We consider the following parameterization for the population distribution \cite{ICAROGW, Mastrogiovanni:2021wsd, Mancarella:2021ecn}:
\begin{equation}\label{eq: pop distr}
    \frac{dN}{d m_1 dm_2 d z}= p(m_1 , m_2 | \bm \Lambda_m) \psi(z|\bm\Lambda_z)\frac{1}{(1+z)}\frac{dV_c(z)}{dz}\,,
\end{equation}
where $\frac{dV_c(z)}{dz}$ is the differential comoving volume per unit of redshift, $p(m_1 , m_2 | \bm \Lambda_m)$ is the distribution of source-frame component masses, and $\psi(z|\bm\Lambda_z)$ is the merger rate density. 

\subsection{Mock data}\label{sec:mock}

To simulate a set of BBH detectable by ET, we will adopt the mass and redshift distributions given by Equations \eqref{eq:mass distrib} and \eqref{eq:redsf distrib}, respectively, and the definition of the cosmological distance given in Equation \eqref{eq: d_L} that encodes the cosmological model. In the Table \ref{tab:inj_pars}, we list the fiducial values of $\bm \Lambda$ parameters,  while the local rate $R_0$ is set to the value inferred from the
GWTC-2 catalog, i.e. $R_0 = 23.9 ^{+14.9}_{-8.6}\ \text{Gpc}^{-3}\text{yr}^{-1} $\cite{LIGOScientific:2020kqk}.
\begin{table}[t!]
    %\centering
    \begin{tabular}{c|c||c|c}
    \hline
    \hline
       \textbf{Parameter}  &  \textbf{Fiducial Value} & \textbf{Parameter}  &  \textbf{Fiducial Value}\\
       \hline
        $\alpha$ & 3.4 & $\delta_m$ & 4.8\\
        $\beta$ & 1.1 & $\kappa$ & 3.0\\
        $m_{min}$ & 5.1 &$z_p$ & 2.0\\
        $m_{max}$ & 87 &$\gamma$ & 2.7\\
        $\mu_{g}$ & 34 & $H_0$& 67.7\\
        $\sigma_{g}$ & 3.6 & $\Omega_{m,0}$ & 0.308\\
        $\lambda_{g}$ & 0.034 & $R_0$ & 23.9 \\
        \hline
        \hline
    \end{tabular}
    \caption{Fiducial values  of astrophysics and cosmological parameters. In particular, the astrophysical parameters are selected in order to be compatible with the constraints inferred from
    the  GWTC-3 \cite{KAGRA:2021duu}, and the cosmological parameters are taken from the latest Planck results \cite{Planck:2018vyg}.}
    \label{tab:inj_pars}
\end{table}

Assuming an observing time $T_{obs}= 1$ yr, we estimate the total number of events by integrating the equation \eqref{eq: pop distr} over the redshift interval and multiplying it by the observing time:
\begin{equation}
    N_{tot}= T_{obs} \int_{0}^{z_{max}} \left[ \frac{dN}{d m_1 dm_2 d z} \right]d m_1 dm_2 d z\,,
\end{equation}
where we have set $z_{max}=15$ to be consistent with the analysis made in \cite{Branchesi:2023mws,Iacovelli:2022mbg}. Using the fiducial values listed in the Table \ref{tab:inj_pars}, we get a total number of BBHs coalescing event of {\bf $\sim10^5$} in one year of observations.  Once the catalog is generated, we focus on the capability of ET for detecting a specific source and analyzing the source  single event parameters. Following \cite{Iacovelli:2022mbg,Dupletsa:2022scg,Ronchini:2022gwk,Muttoni2023,Califano:2024xzt}, we use the signal–to–noise ratio (SNR) as an indicator of  the detectability of a source and the Fisher information matrix (FIM) to estimate the source  single event parameters. To compute the expected SNR and the FIM, we adopt the publicly available code \texttt{GWFISH} \cite{Dupletsa:2022scg}. Here, we briefly recap the main steps. Assuming that one can write the detector output, $s(t)$, as
\begin{equation}
    s(t) = h(t) + n(t)\,,
\end{equation}
where $h(t)$ is the GW signal, and $n(t)$ is detector noise, which is taken to be Gaussian with zero mean and stationary, the GW likelihood is \cite{Cutler:1994ys} 
\begin{equation}
    \label{eq: GW likelihood}
    \mathcal{L}_{GW}(\bm{\theta})=\exp{\{-\frac{1}{2}( s_i - h_i(\bm{\theta})|s_i - h_i(\bm{\theta})) \}} \,,
\end{equation}
where $(\cdot |\cdot)$ is the inner product defined by
\begin{equation}\label{eq: inner prod}
    (a|b) = 4 \Re{\int_{f_{min}}^{f_{max}} \frac{\tilde{a}^*(f) \Tilde{b}(f)}{S_{n}(f)}df}\,.
\end{equation}
In the previous equation, $f_{min}$ and $f_{max}$ are the detector minimum and maximum frequency; $S_n$ is the detector noise power spectral density (PSD)\footnote{We use the publicity available PSD for the ET-D configuration that can be found at \url{https://www.et-gw.eu/index.php/etsensitivities}}; finally, the tilde denotes a temporal Fourier transform.
Starting from equation \eqref{eq: inner prod}, one can define the SNR, $\rho$, as follows
\begin{equation}
    \rho = \sqrt{(h|h)}\,.
\end{equation}
It is worth remembering that since in the ET-D configuration one has three interferometers, the total SNR is given by
\begin{equation}
    \rho_{tot}=\sqrt{\sum_{i=1}^{3}\rho_i}\,.
\end{equation}
In the limit of large SNR, we can expand equation\eqref{eq: GW likelihood} in Taylor series obtaining 
\begin{equation}
    \mathcal{L}_{GW}(\bm{\theta}) = \exp{\{ -\frac{1}{2}\Delta\theta^a\mathcal{F}_{ab}\Delta\theta^b \}},
\end{equation}
where $\Delta\bm{\theta} = \bm{\theta} - \bar{\bm{\theta}}$, with $\bar{\bm{\theta}}$ are the true values of GW parameter, and $\mathcal{F}_{ab}$ is the Fisher matrix defined by
\begin{equation}
    \mathcal{F}_{ab} =\left(\frac{\partial h}{\partial \theta^a}\middle| \frac{\partial h}{\partial \theta^b}\right)_{\theta = \bar{\theta}}\,.
\end{equation}
Once we have computed $\mathcal{F}_{ab}$, we can estimate the uncertainty on the set of parameters $\bm{\theta}$ through the relation: $\sigma_{\theta^a}=\sqrt{(\mathcal{F}^{-1})^{aa}}$.
To generate the synthetic GW signals, we adopted the inspiral-merger-ringdown (IMR) model IMRPhenomXPHM, which includes contributions from both higher-order harmonics and precessing spins~\cite{Pratten:2020ceb}. The true values of $m_1,\ m_2$ and $z$ are sampled from the mass  and redshift distribution. Finally, given the redshift and the value of cosmological parameters, we estimate the luminosity distance $d_L$ through Equation \eqref{eq: d_L}. {We adopt physical prior on the single-event uncertainties. In particular, we impose positive masses and distances, the angles dec, ra, $\psi$ and $\theta_{j_n}$ were set being in the the intervals $[-\pi/2, \pi/2]$, $[0, 2 \pi]$, $[0, 2 \pi]$, $[0,  \pi]$, respectively. Also, we draw uniform spin magnitudes between 0 and 0.99, and isotropic spin directions.
}

Finally, in the Table~\ref{tab: Nevents}, we report the number of detected GW events for different SNR thresholds.
\begin{table}[t]
    \centering
    \renewcommand{\arraystretch}{1.5}
    \begin{tabular}{c|c|c}
    \hline
    \hline
       \textbf{SNR}  &  \textbf{Number of Events} & $\bm{z_{max}}$ \\
       \hline
        200 & 113 & 1.35\\
        150 & 275 & 2.06\\
        100 & 872 & 3.23\\
        80  & 1659 & 6.14\\
        \hline
        \hline
    \end{tabular}
    \caption{Number of BBH events detected by ET at different SNR thresholds, and maximum redshift at which an event is detected ($z_{max}$}.
    \label{tab: Nevents}
\end{table}

\subsection{Hierarchical Bayesian framework}\label{sec:bayesian_frame}

To infer astrophysical and cosmological underlying models jointly, we follow recent works in \cite{Loredo:2004nn,Mandel:2018mve,Thrane_2019, Vitale2020} and use a hierarchical Bayesian approach. This approach is crucial when dealing with GW detections, as the observed sample is inherently biased due to detector sensitivity limitations. By incorporating prior knowledge and population models, hierarchical Bayesian inference enables us to extract the intrinsic distributions of source properties, such as masses and redshifts, and constrain cosmological parameters from GW observations.
The number of detected GW signals follows a Poisson process because GW detections are discrete, stochastic events that occur independently over time, with a mean rate dictated by the underlying astrophysical merger rate and the sensitivity of the detector network. 
Hence, taking into account the Poisson distribution that relates the number of observed events $N_{obs}$ with the expected number of detected events, $N_{exp}$, the hierarchical likelihood can be written as
\begin{equation}\label{eq:hierar_likeli}
\mathcal{L}(\{x\} | \bm \Lambda)\propto \ e^{-N_{exp}(\bm \Lambda)} \prod_{i=1}^{N_{obs}} T_{obs} \int d\bm{\theta}\ \mathcal{L}_{GW}(x_i|\bm{\theta}, \bm \Lambda) \frac{dN}{dtd\bm{\theta}}\,,
\end{equation}
where $\{x\}$ is the dataset associated to the $N_{obs}$ events, $T_{obs}$ is the observing time, $\mathcal{L}_{GW}(x_i|\bm{\theta}, \Lambda)$ is the single event GW likelihood given in equation \eqref{eq: GW likelihood}, and $\frac{dN}{dtd\bm{\theta}}$ is the differential distribution of GW events. Specifically, in order to obtain the rate of GW events we have to integrate $\frac{dN}{dtd\bm{\theta}}$ over all GW parameters $\bm{\theta}$:
\begin{equation}
    \frac{dN}{dt}=\int \frac{dN}{dtd\bm{\theta}} d\bm{\theta}\,.
\end{equation}
Furthermore, the value of the expected number of detected events, $N_{exp}$, accounts for the selection effects according to the definition \cite{Fishbach:2018edt,ICAROGW}:
\begin{equation}\label{eq: Nexp}
N_{exp}(\bm\Lambda) = T_{obs} \int d\bm{\theta} \ p_{det}(\bm{\theta}, \bm\Lambda_{cosmo})\ \frac{dN}{dtd\bm{\theta}}\,.
\end{equation}
The quantity $p_{det}(\bm{\theta},\Lambda)$ is the detection probability of a source characterized by the set $\bm{\theta}$ of GW parameters and $\{\bm{\Lambda} \}$ of  population and cosmological hyper-parameters. The probability of detection is defined by \cite{ICAROGW,Mandel:2018mve, Gair:2022zsa}
\begin{equation}\label{eq: prob det}
    p_{det}(\bm{\theta}, \bm\Lambda_{cosmo}) = \int_{x_{detectable}}dx\ \mathcal{L}_{GW}(x_i|\bm{\theta}, \bm\Lambda_{cosmo})\,.
\end{equation}

The previous probability is, essentially, the integral of GW likelihood over all data, which are detectable, $x_{detectable}$. In our analysis, we consider the SNR threshold for the detection of GW event.
Therefore, using the equation \eqref{eq: Nexp}, the equation \eqref{eq:hierar_likeli} can be recast as \cite{Mastrogiovanni:2021wsd,KAGRA:2021duu}
\begin{equation}
\begin{aligned}\label{eq:hiear_likeli_poisson}
   \mathcal{L}(\{x\} | \bm \Lambda)\propto & \ e^{-N_{exp}(\bm\Lambda)} [N_{exp}(\bm\Lambda)]^{N_{obs}} \times \\
   &\times\prod_{i=1}^{N_{obs}} \frac{\int d\bm{\theta}\ \mathcal{L}_{GW}(x_i|\bm{\theta}, \bm\Lambda) \frac{dN}{dtd\bm{\theta}}}{\int d\bm{\theta} \ p_{det}(\bm{\theta},\bm\Lambda)\ \frac{dN}{dtd\bm{\theta}}}\,. 
   \end{aligned}
\end{equation}

When we neglect the information about the scale $R_0$, 
we assume that the parameters $\bm{\Lambda}$ carry information solely about the shape
of the BBH distribution and that the event rate $R_0$
follows a scale-invariant prior $\propto\frac{1}{R_0}$ \cite{ICAROGW,Mancarella:2025uat}.
Consequently, the equation \eqref{eq:hiear_likeli_poisson} reads \cite{Fishbach:2018edt,Mandel:2018mve}
\begin{equation}\label{eq:MCintegration}
   \mathcal{L}(\{x\} | \bm \Lambda)\propto \prod_{i=1}^{N_{obs}} \frac{\int d\bm{\theta}\ \mathcal{L}_{GW}(x_i|\bm{\theta},  \bm \Lambda) \frac{dN}{dtd\bm{\theta}}}{\int d\bm{\theta} \ p_{det}(\bm{\theta},  \bm \Lambda_{cosmo})\ \frac{dN}{dtd\bm{\theta}}}\,.
\end{equation}

The population distribution is parameterized with respect to the source masses and the redshift, so we have to convert the rate from the detector frame to the source frame \cite{Mastrogiovanni:2021wsd, Mancarella:2021ecn,ICAROGW}
\begin{equation}
    \frac{dN}{dtd\bm{\theta}}=\frac{dN}{dtd m_1^d dm_2^d d d_L}= \frac{1}{J_S^D}\frac{dN}{dtd m_1 dm_2 dz}\,.
\end{equation}
The quantity $J_S^D$ is the Jacobian of the transformation from the detector frame to the source frame and can be expressed as \cite{Mancarella:2021ecn,Mastrogiovanni:2023emh}:
\begin{equation}
    J_S^D = (1 +z)^2 \left(\frac{\partial d_L}{\partial z}\right).
\end{equation}
The derivative of the luminosity distance depends on the underlying of cosmological model and, in our case, is given by 
\begin{align}
\label{eq: dev_d_L}
    \frac{\partial d_L}{\partial z} = \frac{d_L(z)}{1+z}+\frac{c(1+z)}{H_0}\frac{1}{E(z)}\,.
\end{align}
{\color{blue}
Finally, the selection effects, {\em i.e.} the  denominator of Eq. \eqref{eq:MCintegration}, were computed with injections, whose effective number of injections was set to $4N_{obs}$ (\emph{i.e.} $\sim 10^4$) and drawn from the distributions in Eqs. \eqref{eq:mass distrib} and \eqref{eq:redsf distrib}, and re-weighted Monte Carlo integration as in \texttt{ICAROGW} with the posterior samples set to 20 \cite{ICAROGW}. While the condition $m_2<m_1$ is imposed upstream in the process, and checked at several points to ensure it holds.
}

\section{Results}\label{sec:results}

\begin{figure*}[ht!]
    \includegraphics[width=0.99\textwidth]{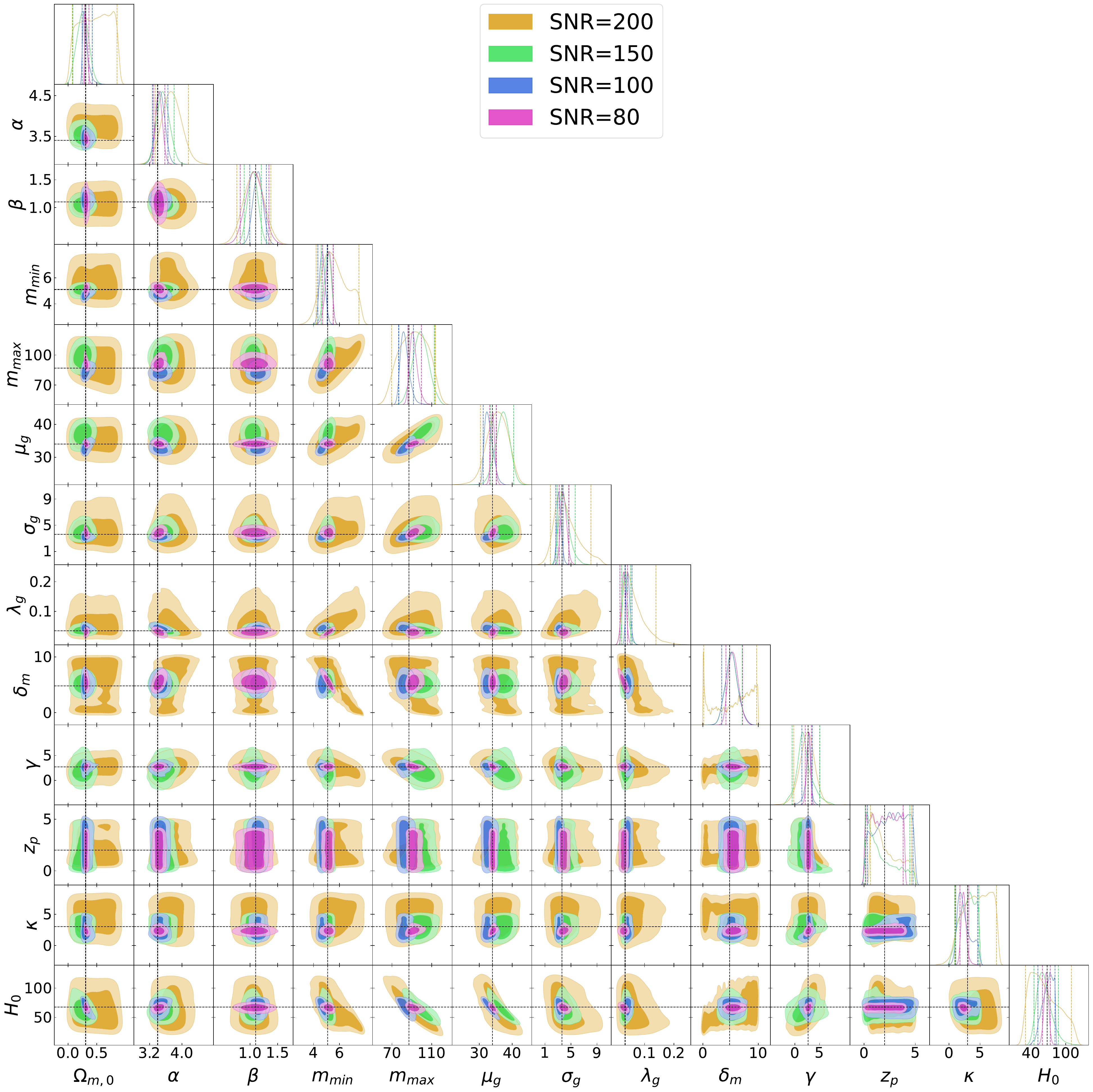}
    \caption{The figure draws the contours of the posterior probability distributions of the astrophysical and cosmological parameters at the 68\% and 95\% credible level. Different colors refer to different SNR thresholds as indicated in the legend. The straight lines indicate the injected fiducial values for the parameters listed in Table \ref{tab:inj_pars}.}
    \label{fig: corner}
\end{figure*}

We have used the four catalogs of GW events listed in the Table \ref{tab: Nevents}, and generated adopting the fiducial model described in Section \ref{sec:theory}  whose parameters were set to the values listed in the Table \ref{tab:inj_pars}, to forecast the precision on mass and redshift distributions and on the cosmological parameters. To achieve such a result, we used \texttt{ICAROGW}, a Python-based tool developed to infer the population distributions of BBHs observed through GWs~\cite{ICAROGW}. Specifically, we employ the \texttt{bilby-mcmc} algorithm to sample the population parameters~\cite{Ashton:2021anp}. The priors used are listed in the Table~\ref{tab:priors}, while we present the results in Table~\ref{tab: result_DS}, and the posterior distributions in Figure~\ref{fig: corner}.
\begin{table}[ht!]
    \centering
    \renewcommand{\arraystretch}{1.5}
    \begin{tabular}{c|c||c|c}
    \hline
       \textbf{Parameter}   &\textbf{Prior}& \textbf{Parameter}  & \textbf{Prior}\\
       \hline
        $\alpha$ & $\mathcal{U}(-3,10)$ &$\delta_m$ & $\mathcal{U}(0,10)$\\
        $\beta$ & $\mathcal{U}(-4,10)$ &$\kappa$ & $\mathcal{U}(-6,6)$\\
        $m_{min}$  &$\mathcal{U}(2,10)$&$z_p$ & $\mathcal{U}(0,5)$\\
        $m_{max}$  &$\mathcal{U}(60,120)$&$\gamma$ & $\mathcal{U}(-4,12)$\\
        $\mu_{g}$  & $\mathcal{U}(20,50)$&$H_0$& $\mathcal{U}(20,120)$\\
        $\sigma_{g}$  &$\mathcal{U}(0.4,10)$ &$\Omega_{m,0}$ &$\mathcal{U}(0.01,0.9)$\\
        $\lambda_{g}$  & $\mathcal{U}(0.0,0.7)$& & \\
        \hline
    \end{tabular}
    \caption{Priors on astrophysical and cosmological parameters.}
    \label{tab:priors}
\end{table}
\begin{table*}[ht!]
    \centering
    \renewcommand{\arraystretch}{1.5}
    \setlength{\tabcolsep}{0.6em}
    \begin{tabular} { c| c| c |c|c|c |c |c|c|c}
    \hline
    \hline
 \textbf{Parameter} & \textbf{Input value} & \textbf{SNR=200} & $\bm{\Delta (\%)}$ & \textbf{SNR=150}& $\bm{\Delta (\%)}$ & \textbf{SNR=100}& $\bm{\Delta (\%)}$ & \textbf{SNR=80} & $\bm{\Delta (\%)}$ \\
\hline
{\boldmath$\Omega_{m,0}   $} & 0.308& $0.48^{+0.37}_{-0.20}      $ & 60 & $0.26\pm 0.10              $ & 38 & $0.31^{+0.02}_{-0.07}   $ & 16 & $0.31^{+0.01}_{-0.03}$ & 6  \\
{\boldmath$H_0            $} & 67.7 & $64^{+20}_{-30}            $ & 39 &  $63^{+9}_{-20}             $ & 24 & $69^{+10}_{-5}             $ & 11 & $66.4\pm 4.2               $ & 6 \\
{\boldmath$\alpha         $} &3.4& $3.75\pm 0.25              $ & 7 & $3.54\pm 0.16              $ & 5 & $3.46\pm 0.12              $ & 4 &  $3.44^{+0.08}_{-0.09}   $ &  3 \\

{\boldmath$\beta          $} &1.1& $1.07\pm 0.19              $ & 18 & $1.05\pm 0.10           $ & 10 & $1.14\pm 0.09            $ & 8 & $1.08\pm 0.16              $ & 15 \\

{\boldmath$m_{min}        $} & 5.1&
$5.70^{+0.78}_{-1.30}       $ & 18 &  $5.08^{+0.30}_{-0.24}      $ & 5 & $4.68^{+0.20}_{-0.23}      $ & 5 & $5.15^{+0.28}_{-0.20}      $ & 4  \\

{\boldmath$m_{max}        $} &87& $92\pm 10                  $ & 11 & $99.0\pm 7.9               $ & 8 & $83.2^{+3.1}_{-5.3}        $ & 5 &  $92.0^{+2.9}_{-4.6}        $ & 4  \\

{\boldmath$\mu_{g}        $} & 34&$35.5^{+3.4}_{-3.0}        $ & 9 & $37.1^{+2.3}_{-2.0}        $ & 6 & $32.91^{+0.89}_{-1.50}      $ & 4 & $34.12^{+0.56}_{-0.64}     $ & 2 \\

{\boldmath$\sigma_g       $} & 3.6&$4.3^{+1.1}_{-2.2}         $ & 40 & $4.03^{+0.57}_{-0.99}      $ & 19 &$3.19^{+0.29}_{-0.38}      $ & 20 & $3.89^{+0.41}_{-0.49}      $ & 12\\

{\boldmath$\lambda_g      $} & 0.034&$0.065^{+0.016}_{-0.043}   $ & 45 & $0.036^{+0.007}_{-0.011} $ & 25 & $0.044^{+0.005}_{-0.008}$ & 16 & $0.029\pm 0.008          $ & 27 \\

{\boldmath$\delta_m       $} & 4.8& $5.5^{+4.5}_{-5.6}         $ & 92 & $5.2\pm 1.1                $ & 21 & $5.2^{+1.1}_{-1.2}         $ & 21 &$5.52^{+0.78}_{-1.00}       $ & 16  \\

{\boldmath$\gamma         $} &2.7& $2.5\pm 1.6                $ & 64 & $2.0^{+1.2}_{-1.7}         $ & 75 & $2.68^{+0.65}_{-0.35}      $ & 19 & $2.75^{+0.39}_{-0.34}      $ & 13 \\

{\boldmath$\kappa         $} & 3.0 &$4.5^{+3.0}_{-1.6}         $ & 51 & $2.99^{+1.70}_{-0.87}       $ & 43 & $2.52^{+0.62}_{-1.4}       $ & 40 & $2.40^{+0.33}_{-0.46}      $ & 16 \\
{\boldmath$z_p            $} & 2.0&$2.2^{+1.3}_{-1.7}         $ & 68 & $1.72^{+0.58}_{-1.70}       $ & 66 & $2.7^{+2.1}_{-1.1}         $ & 59 & $2.0\pm 1.2                $ & 60 \\
\hline
\hline
\end{tabular}
    \caption{The first Column lists the parameters we inferred using the  MCMC algorithm. The second column reports their fiducial values. From the third to the tenth Column, we reported the median value and the 68\% credible level of the posterior distributions for the inferred  astrophysical and cosmological parameters  at each SNR threshold, and relative error on the parameters.}
    \label{tab: result_DS}
\end{table*}

First of all, we want remark that in each run we always recover the fiducial value of each parameter at 68\% of the credible level.
Then, looking at the cosmological parameters, the matter density parameter $\Omega_{m,0}$ shows a factor ten improvement in precision as the SNR decreases from 200 to 80. Using only event with SNR$\geq200$, the accuracy of $\Omega_{m,0}$ is $\sim 59\%$ while, as the threshold decreases to 80, the precision increases dramatically, resulting in a much narrower credible interval and an accuracy of approximately 13.5\%. This trend suggests that data at lower SNRs, possibly due to a higher number of detected events up to redshift $\sim 6$,  provides more reliable constraints on $\Omega_{m,0}$, as expected. Similarly, the Hubble constant $H_0$ exhibits relative accuracies of approximately 39\%, 23\%, 10\%, and 6\% as the SNR threshold decreases from 200 to 80. Although these results still do not reach the precision of neither current analysis based on SNeIa and Cosmic Microwave Background \cite{Planck:2018vyg}, nor future bright siren cosmological analysis \cite{Califano:2022cmo,Califano:2022syd,Califano:2024xzt}, they represent a significant step forward in refining these estimates, in particular considering that we are constraining the cosmological model jointly with the underlying mass and redshift distributions of the BBH population.
\begin{figure}[ht!]
    \centering
    \includegraphics[width=0.48\textwidth]{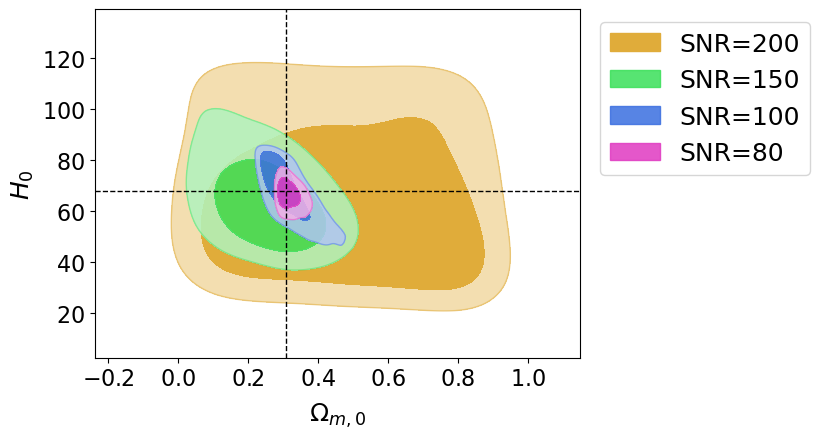}
    \caption{The Figure depicts the contours at the 68\% and 95\% for $\Omega_{m,0}$ and the Hubble constant parameters for the different SNR thresholds considered in our analysis.}
    \label{fig:degeneracy}
\end{figure}
In Figure~\ref{fig:degeneracy}, we illustrate how the degeneracy between the cosmological parameters evolves with different SNR thresholds. 
Using GW event with SNR$\geq200$, the posterior distribution of $\Omega_{m,0}$ is almost uniform. We argue that this is due to both the low number of events and the low redshift at which these events are detected. 
When the SNR threshold decreases to $150$ or $100$, our analysis shows a significant degeneracy between the two cosmological parameters. However, reducing the SNR threshold to 80 allows us to break such a degeneracy and reach a good precision on the cosmological parameters.

Looking at the astrophysical parameters involved in the definitions of the mass distribution of the BBH systems in Equations \eqref{eq:mass_distribution1} and \eqref{eq:mass_distribution2}, the power law indices $\alpha$ and $\beta$ show relative accuracies of approximately 6.6\%, 4.5\%, 3.4\%, and 2.6\% for $\alpha$, and 18\%, 9.5\%, 7.9\%, and 14.8\% for $\beta$, across the SNR thresholds, respectively. The mass distribution is also characterized by the minimum and maximum masses of the binary systems, $m_{min}$ and $m_{max}$, whose relative accuracies are estimated to be of approximately 18.2\%, 5.3\%, 4.7\%, and 4.6\% for $m_{min}$, and 10.8\%, 7.9\%, 5.2\%, and 4.0\% for $m_{max}$, respectively.
The parameters related to the Gaussian component in Equation \eqref{eq:mass_distribution1} of the mass distribution, $\mu_g$ and $\sigma_g$, are constrained with a better precision when the SNR threshold is 80. Thus, we argue that these observations will provide a better resolution for the Gaussian distribution within the mass population achieving a relative accuracy of approximately 1.7\% and 11.6\%, respectively. Additionally, the accuracy on the fraction of events that follows the Gaussian distribution, $\lambda_g$, decreases consistently when SNR decreases from 200 to 80, going from approximately 45\% to 26\%, respectively. Finally, the smoothing parameter $\delta_m$ exhibits distinct correlations with minimum mass of the binary system. Specifically, the increase in $m_{min}$ can be compensated by a simultaneous decrease in $\delta_m$ while still fitting the same population model. 
\begin{figure}[ht!]
    \centering
    \includegraphics[width=0.48\textwidth]{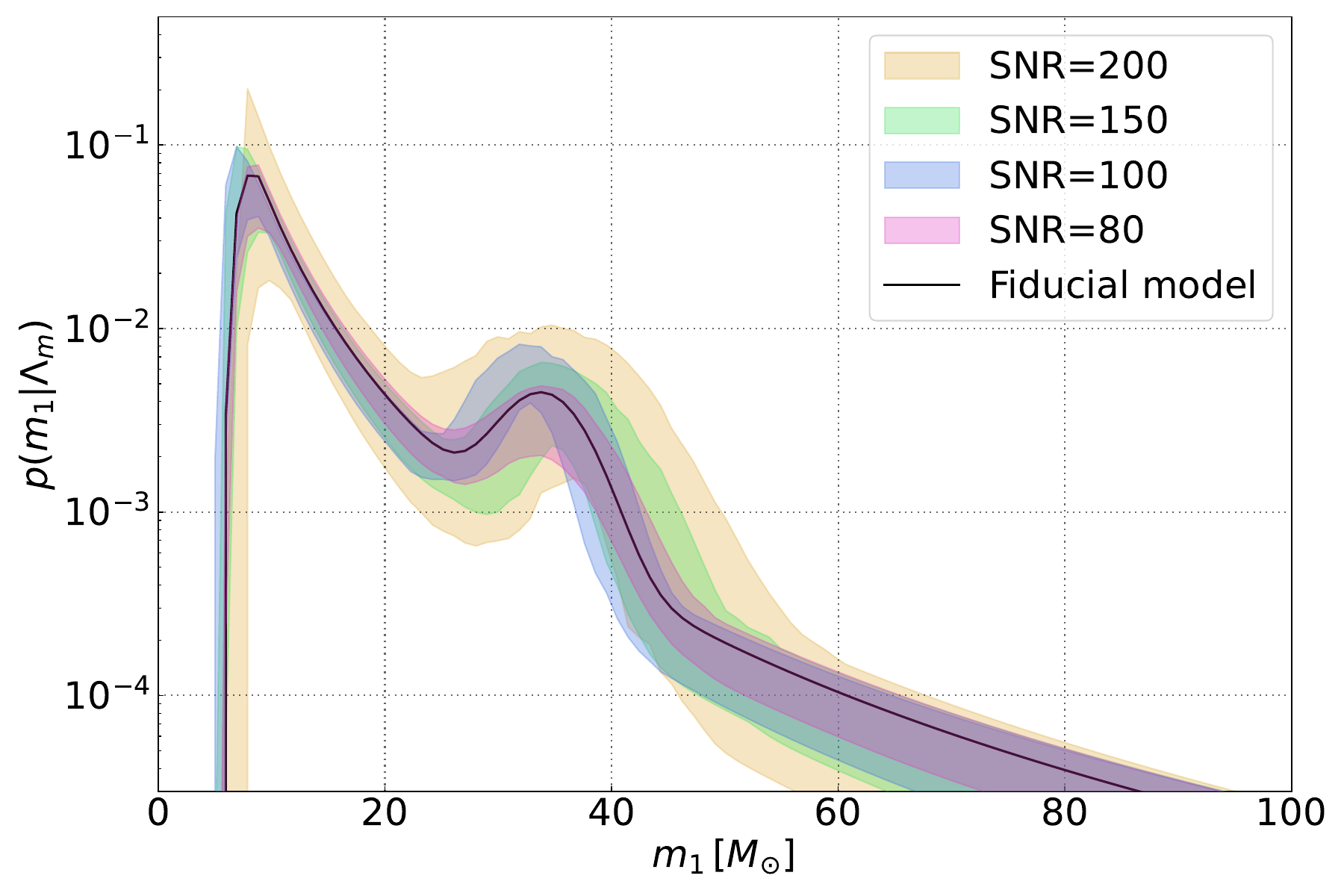}
    \caption{The Figure shows the fiducial mass distribution given in equation \eqref{eq:mass distrib} with parameters listed in Table \ref{tab:inj_pars} (black solid line) and the contours at the 68\% credible level (shaded regions)  for the inferred mass distributions for the different SNR thresholds considered in our analysis.}
    \label{fig:mass_distribution}
\end{figure}
In Figure~\ref{fig:mass_distribution}, we depict the fiducial mass distribution given in equation \eqref{eq:mass distrib} with parameters listed in Table \ref{tab:inj_pars} (black solid line) and the contours at the 68\% credible level (shaded regions)  for the inferred mass distributions for the different SNR thresholds considered in our analysis. As the number of detected events increases (\textit{i.e.} decreasing SNR threshold), the confidence region gets more closely to the fiducial model showing the capability to reconstruct after only one year of observations the mass distributions of BBH systems.
\begin{figure}[ht!]
    \centering
    \includegraphics[width=0.48\textwidth]{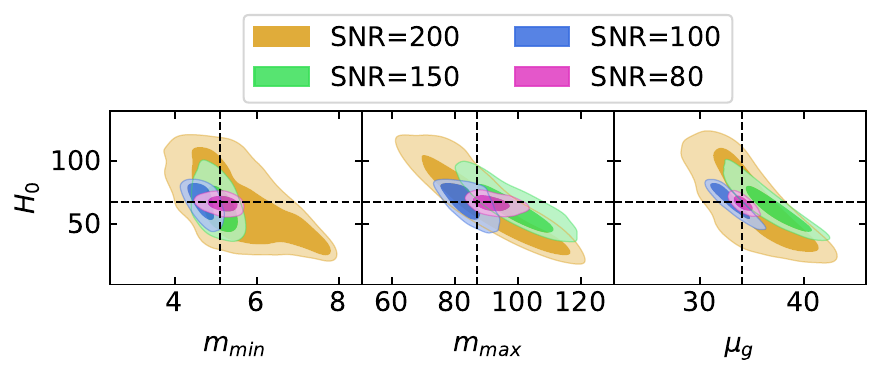}
    \caption{The Figure shows the degeneracies between $H_0$ and some astrophysical parameters related to the mass distribution, namely $m_{min}$, $m_{max}$, and $\mu_g$. }
    \label{fig:degeneracy2}
\end{figure}
Finally, for sake of completeness, in the Figure~\ref{fig:degeneracy2}, we highlight the degeneracy between $H_0$ and some astrophysical parameters related to the mass distribution, namely $m_{min}$, $m_{max}$, and $\mu_g$. The degree of correlation reduces when SNR$=80$ is considered. More importantly, a correlation between the mass distribution and cosmological parameters would prevent constraining them separately without resulting in biased results.

\begin{figure*}[ht!]
    \centering
    \includegraphics[width=0.98\linewidth]{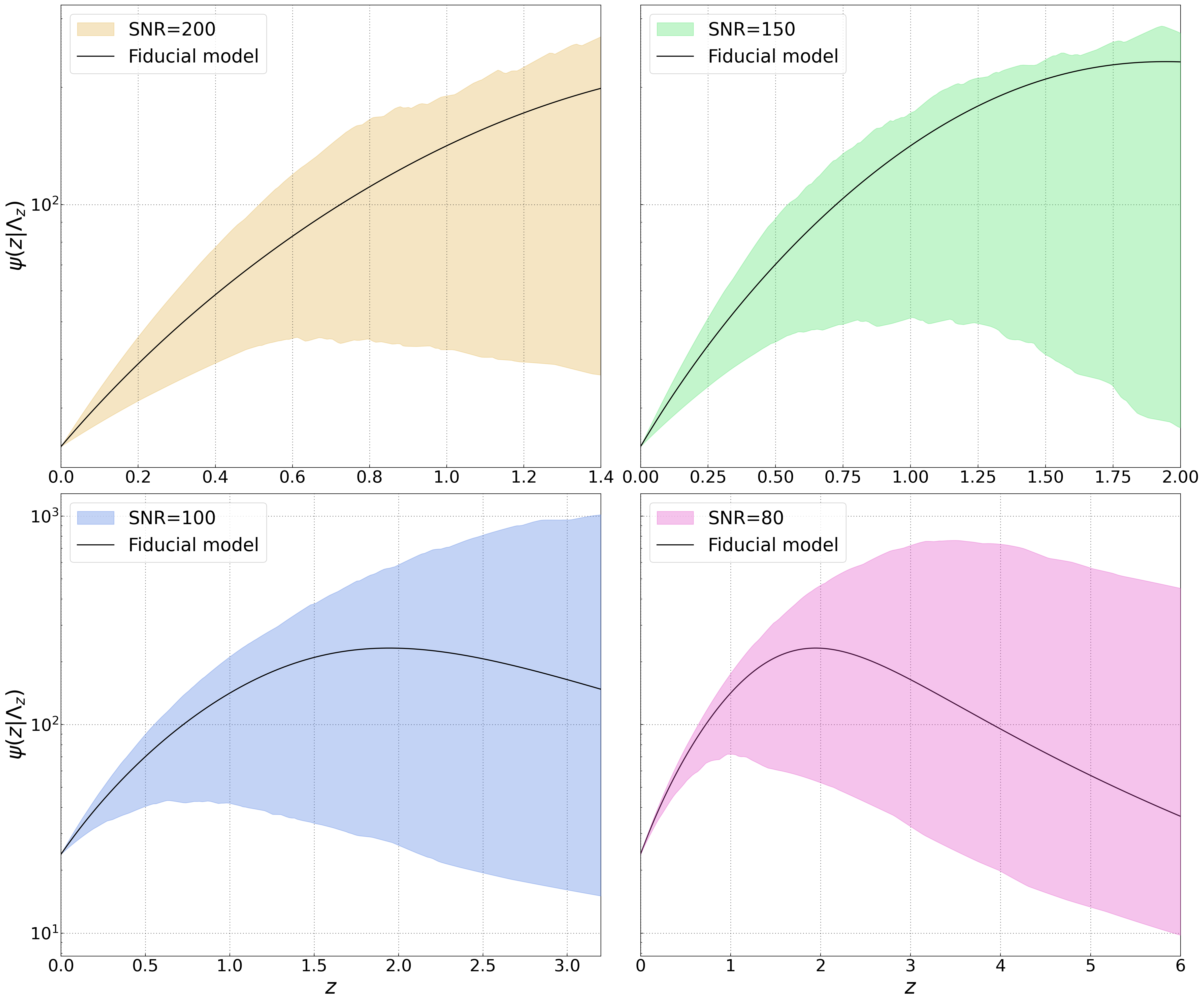}
    \caption{The figure depict the fiducial redshift distribution with parameter set to the values given in Table \ref{tab:inj_pars} (black solid lines), and the contours at the 68\% of the credible level (shaded regions) for the inferred normalized  redshift distribution given in equation \eqref{eq:redsf distrib}. Each panel correspond to a different SNR threshold as reported in the legends, and the redshift evolution stops at the maximum redshift of the mock sample considered.}
    \label{fig:z_distr}
\end{figure*}

Lastly, the astrophysical parameters involved in the definitions of the redshift distribution, $\gamma$, $\kappa$, and $z_p$, are more challenging to constrain accurately. The power to constrain these parameters is closely related to the fiducial maximum redshift of our dataset, and hence to the SNR threshold. In fact, if $z_{max}<z_p$ then the data are totally insensitive to the slope parameter $\kappa$. Nevertheless, constraining $z_p$ proves difficult, the results setting the SNR$\geq80$ are the most stable and precise, particularly for $\gamma$ and $\kappa$, achieving accuracies of roughly $13\%$ and 17\%, respectively, while $z_p$ can be determined with an accuracy of at most $\sim50\%$. In Figure~\ref{fig:z_distr}, we illustrate the fiducial redshift distribution with parameter set to the values given in Table \ref{tab:inj_pars} (black solid lines), and the contours at 68\% of the credible level (shaded regions) for the inferred normalized redshift distribution given in equation \eqref{eq:redsf distrib}. The figure is divided into four separate plots, each corresponding to a specific SNR threshold because varying the SNR threshold alters the maximum redshift that can be reached. As expected, the redshift distribution results poorly constrained with only one year of observations.

\section{Discussion and conclusions}\label{sec:conclusions}

The advent of 3G GWs detectors will cause a paradigm shift, pushing us to move from the analysis of single events from which we usually extract cosmological information, such as the Hubble constant \cite{LIGO_H0_2017}, to the analysis of populations of events which will allow us to constrain both the underlying astrophysical and cosmological models jointly. Here, we have investigated the capability of constraining the mass and redshift distributions 
of BBH systems jointly with the underlying cosmological model using one year of observations of ET. We fix the underlying cosmological model to a flat $\Lambda$CDM model, then we considered the mass distribution given in Equation \eqref{eq:mass distrib}, and the redshift distribution given by the Madau-Dickinson model in Equation \eqref{eq:redsf distrib}. The fiducial values of the model parameters are listed in Table \ref{tab:inj_pars}. Then, we built mock catalogs  mock catalogs with different SNR thresholds, and finally inferred astrophysical and cosmological parameter jointly adopting a  hierarchical Bayesian framework. 

The analysis considered four SNR thresholds, specifically 200, 150, 100 and 80.
In particular, we have shown that as the SNR threshold decreases, the precision on cosmological parameters, \textit{i.e.} the matter density parameter $\Omega_{m,0}$ and the Hubble constant $H_0$, improves significantly. We argue this is likely due to the increased number of detectable events at high redshift when setting lower SNR thresholds (as shown in the Table \ref{tab: Nevents}), which provides a richer dataset and enables more reliable constraints. However, \textit{dark sirens} still face challenges in achieving competitive precision compared to bright sirens, as demonstrated by the broader uncertainties in our forecasts (see Table \ref{tab: result_DS}).
We also examined how the degeneracy between cosmological and astrophysical parameters evolves as we vary the SNR thresholds. At higher SNR thresholds, a notable degeneracy exists between $\Omega_{m,0}$ and $H_0$, which is only mitigated at lower SNRs where the precision of the estimates increases. Furthermore, our analysis of the mass and redshift distributions showed that the constraints on astrophysical parameters, such as the minimum and maximum black hole masses and the parameters governing the redshift distribution, benefits from a higher number of detected events. We also have shown that one year of observations will be enough to reconstruct the mass distribution with its features. Conversely, the redshift distribution is poorly constrained and will need more observations to get a better precision. This will be of enormous importance if one wants to test the possible existence of a subpopulation of primordial black holes whose redshift distribution behaves differently at high redshift.

Our analysis is based on certain assumptions that affect the final results. Here, we discuss these assumptions in more detail. First, we assume that the mass distribution is not evolving with redshift. Our hypothesis relies on analysis of the GWTC-2 events, where the mass distribution model does not show a redshift evolution \cite{Fishbach:2021yvy}. In fact, any redshift-dependent variations are expected to be mild, particularly within the redshift range covered by GWTC-3 \cite{Mapelli:2019bnp}. Although no evidence for redshift dependence has been observed within the range probed by the LVK Collaboration, such an assumption might be invalidated by 3G detectors \cite{Fishbach:2021yvy,vanSon:2021zpk,Karathanasis:2022rtr,Belczynski:2022wky}, which will reach higher redshifts. Consequently, the mass distribution parameters would evolve with redshift and the degeneracies with cosmological parameters could become more serious \cite{Agarwal:2024hld}. 

Second, we did not take spin distributions into account in our analysis. Neglecting the spin distribution is a simplifying assumption that affects our results. Future studies should aim to incorporate the spin distributions to make these results more realistic, though other analyses have already shown that constraining the spin distribution is challenging \cite{DeRenzis:2024dvx}. 

Third, we adopt a hierarchical Bayesian framework; however, this is known to introduce potential sources of systematic bias \cite{Pierra:2023deu}. The use of inaccurate mass and redshift distributions can lead to biases in cosmological inference. The main sources of bias stem from the model of the mass distribution, which fails to account for redshift evolution and cannot capture unexpected features in the mass spectrum. To address these issues, a non-parametric model for the mass distribution of compact binaries has been developed that successfully reconstructs both the source mass model and cosmological parameters, thus avoiding such biases \cite{Farah:2024xub} {though the mass distribution is not allowed to evolve with respect to the redshift and is therefore still affected by systematic biases.} Future works should consider adopting this model and studying the correlations with cosmological parameters. 

Fourth, we restricted the analysis to events with SNR thresholds greater than 80. Limiting the analysis to events with SNR above 80 ensures higher-quality detections; however, this restriction reduces the number of events considered and potentially impacts the generalizability of our results. Today, the LVK collaboration uses 47 GW sources from the GWTC–3, with an SNR greater than 11, to estimate the Hubble constant. 

Fifth, a Fisher matrix analysis has recently been implemented to carry out population studies \cite{DeRenzis:2024dvx}. The main difference from our approach is that the underlying cosmological model is fixed to study constraints on population parameters alone. {Since cosmological parameters are not inferred from the posterior distributions possible degeneracies with mass and redshift distribution parameters that we have found (see Figure \ref{fig:degeneracy} did not appear in the former study.} Another important difference is that, in \cite{DeRenzis:2024dvx}, the spin distribution is included and inferred. Their main result aligns with our findings that is the capability of 3G detectors in constraining the mass distributions with a good accuracy within one year of observations. 

{Sixth, the detection probability in Eq. \eqref{eq: prob det} should enter the integral over $\theta$ \cite{Essick:2023upv},
otherwise biases could be introduced in the methodology. Nevertheless, under the assumption made in this work, the previous simplification works correctly.}

{Least but not last, in \cite{Mancarella:2025uat},  a full sampling of the hierarchical population posterior distribution of BBH merging was carried out using current GWTC-3 LVK catalog demonstrating that the sampling the full hierarchical problem is feasible, and  the uncertainties associated with some of
the Monte Carlo integrations can be eliminated allowing to converge to results obtained adopting a lower dimensional parameter space and pointing out correlation between parameters.}

{To sum up, we demonstrated the enormous capabilities of ET to constrain population hyperparameters and the underlying cosmological model at the same time after only one year of observing time even when the SNR detection thresholds considered are very high and so there are not many events included in the catalog with respect to the total number of expected detection. }

\acknowledgments

M.C. and D.V. acknowledge the support of Istituto Nazionale di Fisica Nucleare (INFN), {\it iniziative specifiche} MOONLIGHT and TEONGRAV. 
D.V. acknowledges the FCT project with ref. number PTDC/FIS-AST/0054/2021. I.D.M. acknowledges financial support from the grant PID2021-122938NB-I00 funded by MCIN/AEI/10.13039/501100011033 and from the grant SA097P24 funded by Junta de Castilla y León and by “ERDF A way of making Europe”.
This work was realized thanks to computational resources acquired by funds from Project code PIR01\_00011 “IBISCo”, PON 2014-2020.

\bibliography{Biblio}

\end{document}